\documentclass[reprint,showpacs,amsmath,amssymb,aps]{revtex4-1}
\usepackage{graphicx}
\usepackage{bm}
\pdfoutput=1
\begin{document}
\newcommand{\balpha}{\mbox{\boldmath$\alpha$}}

\title{Coulomb and quenching effects in small nanoparticle-based spasers}

\author{Vitaliy N. Pustovit and Augustine M. Urbas}
\affiliation{Materials and Manufacturing Directorate, Air Force Research Laboratory, Wright Patterson Air Force Base, Ohio 45433, USA}

\author{Arkadi  Chipouline}
\affiliation{Institute for Microelectrotechnics and Photonics,
Technical University of Darmstadt, Merckstr. 25, 64283 Darmstadt, Germany}

\author{Tigran V. Shahbazyan}
\affiliation{Department of Physics, Jackson State University, Jackson MS 39217 USA}


\begin{abstract}
  We study numerically the effect of mode mixing and direct dipole-dipole interactions between gain molecules on spasing in a small composite nanoparticles with a metallic core and a dye-doped dielectric shell. By combining Maxwell-Bloch equations with Green's function formalism, we calculate lasing frequency and threshold population inversion for various gain densities in the shell. We find that  gain coupling to nonresonant plasmon modes has a negligible effect on spasing threshold. In contrast, the direct dipole-dipole coupling, by causing random shifts of gain molecules' excitation frequencies, hinders reaching the spasing threshold in small systems. We identify a region of parameter space in which spasing can occur considering these effects.
\end{abstract}


\maketitle

\section{Introduction}
\label{sec:intro}
The prediction of plasmonic laser (spaser) \cite{bergman-prl03,stockman-natphot08,stockman-jo10} and its  experimental realization in various systems \cite{noginov-nature09,zhang-nature09,zheludev-oe09,zhang-natmat10,ning-prb12,gwo-science12,shalaev-nl13,odom-natnano13,gwo-nl14,zhang-natnano14,odom-natnano15,zharov-15} have been among the highlights in the rapidly developing field of plasmonics during the past decade \cite{shahbazyan-stockman}. First observed in gold nanoparticles (NP) coated by dye-doped dielectric shells \cite{noginov-nature09}, spasing action was reported in hybrid plasmonic waveguides \cite{zhang-nature09}, semiconductor quantum dots on metal film \cite{zheludev-oe09,gwo-nl14},  plasmonic nanocavities and nanocavity arrays \cite{zhang-natmat10,ning-prb12,gwo-science12,odom-natnano13,zhang-natnano14,odom-natnano15},  metallic NP and nanorods \cite{shalaev-nl13,zharov-15}, and recently was studied in graphene-based structures \cite{apalkov-light14}. The small spaser size well below the diffraction limit gives rise to numerous promising applications, e.g., in sensing \cite{zhang-natnano14} or medical diagnostics\cite{zharov-15}. However, most experimental realizations of spaser-based nanolasers were carried in relatively large systems, while only a handful of experiments reported  spasing action in small systems with overall size below 50 nm \cite{noginov-nature09,zharov-15}.

The spaser feedback mechanism is based on  near field coupling between   gain and plasmon mode which, in the single-mode approximation, leads to a lasing threshold condition \cite{stockman-jo10},
\begin{equation}
\label{condition}
\frac{\mu^{2}\tau_{2} }{\hbar V_{m}}\,NQ\sim 1.
\end{equation}
where $\mu$ and $\tau_{2}$ are, respectively, the gain dipole matrix element and relaxation time, $N$ is the population inversion, $Q$ is the plasmon mode quality factor, and $V_{m}$ is the mode volume. While Eq. \ref{condition} represents the standard threshold condition  for gain coupled to a resonance mode \cite{haken},  there is an issue of whether this condition needs to be modified in realistic plasmonic systems \cite{stockman-prl11,khurgin-oe12,stockman-prl13,klar-bjn13}. For example, it has long been known that  fluorescence of a molecule placed sufficiently close to a metal surface is quenched due to the Ohmic losses in the metal \cite{silbey-acp78,metiu-pss84}. During past decade, numerous experiments \cite{feldmann-prl02,lakowicz-jf02,artemyev-nl02,gueroui-prl04,feldmann-nl05,klimov-jacs06,strose-jacs06,mertens-nl06,pompa-naturenanotech06,novotny-prl06,sandoghdar-prl06,novotny-oe07,sandoghdar-nl07,lakowicz-jacs07,chen-nl07,lakowicz-nl07,halas-nl07,feldmann-nl08,halas-acsnano09,ming-nl09,kinkhabwala-naturephot09,viste-acsnano10,lakowicz-jacs10,munechika-nl10,ming-nl11,ratchford-nl11,raino-acsnano11} reported fluorescence enhancement by resonant dipole surface plasmon mode in spherical metal NP that was followed by quenching due to coupling to nonresonant modes  as the molecules moved closer to the NP surface \cite{nitzan-jcp81,ruppin-jcp82,pustovit-jcp12}. Another important factor is the direct dipole-dipole interactions between gain molecules which causes random Coulomb shifts of molecules' excitation energies and therefore could lead to the system dephasing \cite{eberly-pra71,friedberg-pra74,friedberg-pra78}.

In this paper, we perform a numerical study of the role of quenching and direct interactions between gain molecules in reaching the lasing threshold for small spherical NP with metal core and dye-doped dielectric shell. We use  a semiclassical approach that combines Maxwell-Bloch equations with the Green function formalism to  derive the threshold condition in terms of exact system eigenstates, which we find numerically. We show that for a large number of gain molecules  needed to satisfy Eq. (\ref{condition}),  the coupling to nonresonant modes plays no significant role. In contrast, the direct dipole-dipole interactions, by causing random shifts in gain molecules' excitation energies,  can hinder reaching the lasing threshold in small NP-based spasers.

The paper is organized as follows. In the first section we describe our model and derive the lasing threshold condition in terms of exact system eigenstates. In the second we present the results of our numerical calculations, and then we conclude the paper.

\section{The model}
\label{sec:model}
\begin{figure}[bt]
 \centering
 \includegraphics[width=0.8\columnwidth]{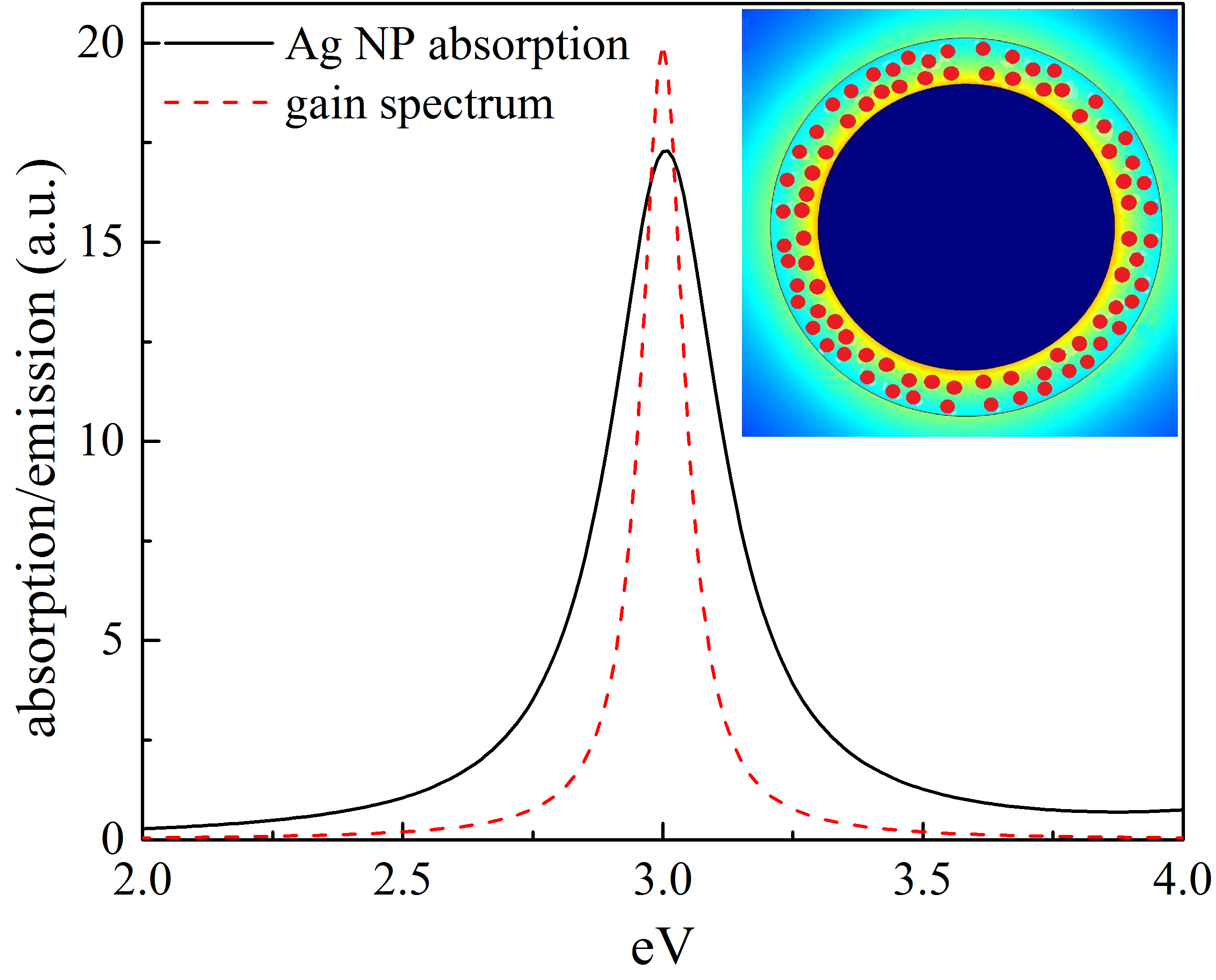}
 \caption{\label{fig1} Normalized  spectra for spherical Ag NP with radius $R=5$ nm and  gain molecule with maximum tuned to plasmon resonance. Inset: Schematics of a composite NP with Ag core and dielectric shell doped with M dye molecules.}
  \end{figure}

We consider a composite spherical nanoparticle (NP) with a metallic core of radius $R_c$ and dielectric shell of thickness $h$ that is doped with $M$ fluorescent dye molecules at random positions $\textbf{r}_{j}$ [see inset  Fig.\ \ref{fig1}]. Within  the semiclassical  approach,  the gain molecules are described by pumped two-level systems, with excitation frequency $\omega_{21}$ between the  lower level 1 and upper level 2, while  electromagnetic fields are treated classically. Each molecule  is characterized by the polarization $\rho_{j}\equiv \rho_{12}^{(j)}$ and population inversion $n_{j}\equiv\rho_{22}^{(j)}-\rho_{11}^{(j)}$, where $\rho_{ab}^{(j)}$ ($a,b=1,2$) is the density matrix for $j$th molecule. In the rotating wave approximation, the steady state molecule dynamics is described by optical Bloch equations \cite{chipouline-jo12}

%
\begin{align}
\label{maxwell-bloch}
& \left[i + \tau_{2}(\omega - \omega_{21}) \right] p_{j} = \frac{\tau_{2}}{\hbar} A_{j}\, n_{j}
\\
&n_{j}-n_0 = \frac{2i\tau_1}{\hbar}\left( A_{j}p_{j}^{*}- A_{j}^{*} p_{j}\right),
\nonumber\\
&n_0=\frac{W \tilde\tau_1 -1}{W \tilde \tau_1 +1},~~~~\tau_1=\frac{\tilde \tau_1}{W \tilde \tau_1 +1},
\nonumber
\end{align}
where $\tau_2$ and  $\tilde{\tau_1}$ are the time constants describing phase and energy relaxation processes,  $W$ is the phenomenological pump rate, and $A_{j}=\mu\textbf{e}_{j}\cdot \textbf{E}(\textbf{r}_{j})$ is the interaction. Here $\textbf{E}(\textbf{r}_{j})$ is the slow amplitude of the local field at the point of $j$th molecule, and $\mu$ and $\textbf{e}_{j}$ are, respectively, the molecule dipole matrix element and orientation. The local field  $\textbf{E}(\textbf{r})$ is created by all molecular dipoles in the presence of a NP and satisfies the Maxwell's equation
\begin{align}
\label{maxwell}
{\bm \nabla} \times {\bm \nabla} \times {\bf E}({\bf r})
&-\epsilon ({\bf r},\omega )\frac{\omega^{2}}{c^2} {\bf E}({\bf r})=\frac{4\pi \omega^{2}}{c^2}\sum_{j}\textbf{p}_{j}\delta(\textbf{r}-\textbf{r}_{j}),
\end{align}
where $\epsilon({\bf r},\omega )$ is the local dielectric function given by metal, shell and outside dielectric functions in the corresponding regions, $c$ is the speed of light, and $\textbf{p}_{j}=\mu\textbf{e}_{j}\rho_{j}$ is the molecule dipole moment. The solution of Eq.~(\ref{maxwell}) has the form
\begin{eqnarray}
\label{electric}
 \textbf{E}({\bf r} ) = \textbf{E}_{0}({\bf r} ) + \frac{4\pi \omega^{2}\mu}{c^2} \sum_{j=1}^{M} \textbf{G}({\bf r},{\bf r}_{j}) \cdot \textbf{e}_{j}\,\rho_{j} ,
\end{eqnarray}
where ${\bf E}_{0}({\bf r})$ is a solution of the homogeneous part of  Eq.~(\ref{maxwell}) (i.e., in the absence of molecules) and  $\textbf{G}({\bf r},{\bf r}')$  is the Green dyadic in the presence of a NP. After expressing the polarization in terms of local fields using Eq. (\ref{maxwell-bloch}),
and then eliminating the local fields using Eq. (\ref{electric}), the system (\ref{maxwell-bloch}) takes the form
\begin{align}
\label{maxwell-bloch2}
&\sum_{k=1}^{M}\left [\left (\omega-\omega_{12}+i/\tau_{2}\right )\delta_{jk}- n_{j} D_{jk}\right ]\rho_{k}= \mu\textbf{e}_{j}\cdot \textbf{E}_{0}(\textbf{r}_{j}),
\nonumber\\
&n_{j}-n_0 +4\tau_1\text{Im}\sum_{k=1}^{M}\left[ \rho_{j}^{*}D_{jk}\rho_{k}\right]=\frac{4\tau_1}{\hbar}\text{Im}\left[\rho_{j}\mu\textbf{e}_{j}\cdot \textbf{E}_{0}^{*}(\textbf{r}_{j})\right],
\end{align}
where $\delta_{jk}$ and  $D_{jk}(\omega)$ are, respectively,  the Kronecker symbol and frequency-dependent coupling matrix in the configuration space given  by
\begin{equation}
\label{matrix}
D_{jk}=\dfrac{4\pi\omega^{2}\mu^{2}}{c^{2}\hbar}\,\textbf{e}_{j}\cdot  \textbf{G}({\bf r}_{j},{\bf r}_{k}) \cdot \textbf{e}_{k}.
\end{equation}
Equations (\ref{maxwell-bloch2}) and (\ref{matrix}) constitute our model for active molecules near a plasmonic NP. For a sufficiently high pump rate, $W\tilde{\tau}_{1}>1$ [see Eq. (\ref{maxwell-bloch})], spasing action is possible  provided that  losses are compensated \cite{bergman-prl03,stockman-natphot08,stockman-jo10}.

We are interested in the collective system eigenstates  defined by the homogeneous part of system (\ref{maxwell-bloch2}),
\begin{align}
\label{maxwell-bloch-hom}
&\sum_{k=1}^{M}\left [\left (\omega-\omega_{12}+i/\tau_{2}\right )\delta_{jk}- n_{j}D_{jk}\right ]\rho_{k}=0,
\nonumber\\
&n_{j}-n_0 + 4\tau_1\text{Im}\sum_{k=1}^{M}\left[ \rho_{j}^{*}D_{jk}\rho_{k}\right]=0.
\end{align}
Following the procedure employed previously for studying plasmon-mediated cooperative emission \cite{pustovit-prl09,pustovit-prb10}, we introduce eigenstates $|J\rangle$ of the coupling matrix $\hat{D}$ as
\begin{equation}
\hat{D}|J\rangle=\Lambda_{J}|J\rangle,
~~~
\Lambda'_{J}+i\Lambda''_{J},
\end{equation}
where $\Lambda'_{J}$ and $\Lambda''_{J}$ are, respectively,  real and imaginary parts of system eigenvalues  $\Lambda_{J}$ which represent the frequency shift and decay rate of an eigenstate $|J\rangle$. We now introduce \textit{collective} variables for polarization and population inversion   as
\begin{equation}
\rho_{J}=\sum_{j}\langle \bar{J}|j\rangle \rho_{j},
~~
n_{JJ'}=\sum_{j}\langle \bar{J}|j\rangle n_{j}\langle j|J'\rangle,
\end{equation}
where, to ensure the orthonormality, we used the eigenstates $|\bar{J}\rangle$ of complex conjugate matrix $\bar{D}_{jk}$ corresponding to the advanced Green function of Eq. (\ref{maxwell}). Multiplying the first equation of system (\ref{maxwell-bloch-hom}) by $\langle \bar{J}|j\rangle $ and then summing both equations over $j$, the system (\ref{maxwell-bloch-hom}) in the  basis of collective eigenstates takes the form
\begin{align}
\label{maxwell-bloch-hom-coll}
&\sum_{J'=1}^{M}\left [\left (\omega-\omega_{21}+i/\tau_{2}\right )\delta_{JJ'}- n_{JJ'}\Lambda_{J'}\right ]\rho_{J'}=0,
\nonumber\\
&N_{0}-N +4\tau_1\sum_{J=1}^{M}\Lambda''_{J} |\rho_{J}|^{2}=0,
\end{align}
where $N=\sum_{j}n_{j}$ is the ensemble population inversion and $N_{0}=n_{0}M$. The mixing of collective states $J$ through $n_{JJ'}$ originates from the inhomogeneity of $n_{j}$ distribution for individual molecules. In the following, we assume that, for a sufficiently large ansemble, this inhomogeneity is weak  and adopt $n_{JJ'}=n\delta_{JJ'}$, where $n=N/M$ is the \textit{average} population inversion per molecule. Note that, in this approximation,  the individual molecule polarizations $\rho_{j}$ are still random due to the molecules' spatial distribution. The first equation of system (\ref{maxwell-bloch-hom-coll}) then yields the characteristic equation for each state,
\begin{align}
\label{maxwell-bloch-hom-coll1}
\omega-\omega_{21}+i/\tau_{2}-n\Lambda_{J}(\omega)=0,
\end{align}
implying that each eigenstate acquires self-energy $n\Lambda_{J}(\omega)$ due to the interactions of molecules with the  NP and each other. The resonance frequency of mode $J$ is determined by the real part of Eq. (\ref{maxwell-bloch-hom-coll1}),
\begin{align}
\label{resonance}
\omega= \omega_{21} + n\Lambda'_{J} (\omega) ,
\end{align}
while its imaginary part,
\begin{equation}
\label{loss_condition}
 n\tau_{2}\Lambda''_{J}(\omega) =1,
\end{equation}
determines $n$ and, in fact, represents the lasing threshold condition.
Eliminating $n$, we obtain the equation for resonance frequency $\omega$,
\begin{eqnarray}
\label{resonance2}
\tau_2\left (\omega-\omega_{21}\right ) = \Lambda'_{J}(\omega)/\Lambda''_{J}(\omega).
\end{eqnarray}
Equations (\ref{resonance})-(\ref{resonance2}) are valid for any plasmonic system with weak inhomogeneity of gain population inversion. For a spherical core-shell NP that we consider, the plasmon modes are characterized by angular momentum $l$ and by well-separated frequencies  $\omega_{l}$. However, each system eigenstate $|J\rangle$ contains, in general, contributions from all $l$ since NP spherical symmetry is broken down by the random distribution of molecules within the shell.

In order to establish the relation of our model to a conventional spaser description \cite{bergman-prl03,stockman-natphot08,stockman-jo10}, let us assume for now a largely homogeneous spatial distribution of molecules in the shell and disregard the effects of direct dipole-dipole interactions. This could be considered one extreme of real systems where dyes do not interact due to mutual orientation and distribution. In this case,  the eigenstates $|J\rangle$ are dominated by molecules' coupling with the $l$th plasmon mode and can be labeled as $\Lambda_{l}$. Assume now that  gain excitation energy is close to some $l$th plasmon energy, $\omega_{21}\approx \omega_{l}$. In this case, for small overall system size, there is a ($2l+1$)-fold degenerate eigenstate of matrix (\ref{matrix}) which scales linearly with the number of molecules as
\begin{equation}
\label{coop}
\Lambda_{J}\sim M\lambda_{l},
\end{equation}
where $\lambda_{l}$ is the \textit{single molecule} self-energy \cite{pustovit-prl09,pustovit-prb10,josab}. For example, the single-molecule self-energy $\lambda_{1}$ due to the near-field coupling to  dipole ($l=1$) plasmon mode  is given by \cite{pustovit-prl09,pustovit-prb10,josab}  (also see below)
\begin{align}
\lambda_{1}= \frac{4\mu^{2}}{\hbar}\frac{\alpha_{1}(\omega)}{r^{6}},
\end{align}
where $r$ is the average distance to   NP center and $\alpha_{1}(\omega)$ is NP dipole polarizability (for simplicity, we assumed normal dipole orientation relative to the NP surface). Near the plasmon resonance $\omega\sim\omega_{p}$, the NP polarizability can be approximated as
\begin{align}
\label{alpha1}
\alpha_{1}(\omega)\sim \frac{R^{3} \omega_{p}}{\omega_{p}-\omega-i/\tau_{p}},
\end{align}
where  $\tau_{p}$ is the plasmon lifetime and $R$ is the overall NP size. Then Eq. (\ref{resonance2}) yields the standard expression for resonance frequency \cite{bergman-prl03,stockman-natphot08,stockman-jo10}
\begin{equation}
\label{resonance3}
\omega_{s}=\frac{\omega_{p}\tau_{p}+\tau_{2}\omega_{21}}{\tau_{p}+\tau_{2}}.
\end{equation}
For exact molecule-plasmon resonance, $\omega_{21}=\omega_{p}$, the solution of Eq. (\ref{resonance2}) is $\omega=\omega_{21}=\omega_{p}$ (i.e., there is no frequency shift), and we have $\alpha''_{1}\sim R^{3}Q$, where $Q=\omega_{p}\tau_{p}$ is the plasmon quality factor. Then, for $r\sim R$,  Eq.  (\ref{loss_condition}) takes the form
\begin{equation}
\label{condition1}
\frac{\mu^{2}\tau_{2}}{\hbar R^{3}}\, NQ\sim 1,
\end{equation}
where we used $N=nM$. Since for small NPs, the local fields penetrate  the entire system volume, i.e., $V_{m}\sim R^{3}$, the conditions  (\ref{condition1}) and (\ref{condition}) coincide.

For general gain  distribution in the shell, each of the exact system eigenstates contains a contribution  from nonresonant plasmon modes. For a single fluorescing molecule coupled to a dipole plasmon mode, the high $l$ modes' contribution  leads to fluorescence quenching if the molecule is sufficiently close to the metal surface \cite{silbey-acp78,metiu-pss84,feldmann-prl02,lakowicz-jf02,artemyev-nl02,gueroui-prl04,feldmann-nl05,klimov-jacs06,strose-jacs06,mertens-nl06,pompa-naturenanotech06,novotny-prl06,sandoghdar-prl06,novotny-oe07,sandoghdar-nl07,lakowicz-jacs07,chen-nl07,lakowicz-nl07,halas-nl07,feldmann-nl08,halas-acsnano09,ming-nl09,kinkhabwala-naturephot09,viste-acsnano10,lakowicz-jacs10,munechika-nl10,ming-nl11,ratchford-nl11,raino-acsnano11}. At the same time, the role of direct dipole-dipole interactions between gain molecules confined in a small volume may be significant as well due to large Coulomb shifts of molecules' excitation frequencies \cite{eberly-pra71,friedberg-pra74,friedberg-pra78}. Both the mode-mixing and direct coupling effects can be incorporated on an equal footing within our approach through the corresponding terms in the matrix (\ref{matrix}). The results of our numerical calculations are presented in the next section.

\section{Numerical Results and Discussion}
\label{sec:num}
Numerical calculations were carried out for ensembles of $M=600$ and $M=1000$ dye molecules randomly distributed within a Silica shell of uniform thickness $h$ in the range from 0.5 to 3 nm on top of a spherical Ag NP of radius $R_{c}=5$ nm.  Note that thicker shells pose numerical challenges as they require a significantly larger number of gain molecules to satisfy Eq. (\ref{condition}). For the same reason, we assume a normal orientation of molecules' dipole moments relative to the NP surface. In this case, the matrix (\ref{matrix}) takes the form $D_{jk}=D_{jk}^{d}+D_{jk}^{p}$, where $D_{jk}^{d}$ and $D_{jk}^{p}$ are, respectively, the direct (dipole-dipole) and plasmonic contributions given by \cite{pustovit-prb10}
\begin{align}
\label{matrix1}
 &D_{jk}^{d}(\omega)=-(1-\delta_{jk})
 \frac{\mu^{2}}{\hbar} \frac{\varphi_{jk}}{ r^3_{ij}},
\nonumber\\
&D_{jk}^{p}(\omega)= \frac{\mu^{2}}{\hbar}\sum_l \alpha_l (\omega)(l+1)^2  \frac{ P_l (\cos
\gamma_{jk}) } { r_i^{l+2} r_j^{l+2}},
\end{align}
where  $\alpha_{l}$ is the NP $l$th multipolar polarizability, $P_l (\cos\gamma_{jk})$ is the Legendre polynomial of order $l$,  $\gamma_{jk}$ is the angle between molecule locations ${\bf r}_j$ and ${\bf r}_k$, $\varphi_{jk}=1+ \sin^2 (\gamma_{jk}/2)$ is the orientational factor in the dipole-dipole interaction term, and $r_{jk}=|{\bf r}_j-{\bf r}_k|$. The gain frequency $\omega_{21}$ was tuned to the $l=1$ plasmon frequency $\omega_{p}=3$ eV (see Fig. \ref{fig1}) and its bandwidth and dipole matrix element were taken as $\hbar/\tau_{2}=0.05$ eV and $\mu=4$ D, which are typical values for the Rhodamine family of dyes. The NP was embedded in a medium with dielectric constant $\epsilon_m = 2.2$ and we used the Drude form of Ag dielectric function \cite{coronado, christy} for calculation of NP polarizabilities, while the plasmon damping rate was appropriately modified to incorporate Landau damping in a small NP.

The eigenstates were found by numerical diagonalization of matrix $D_{jk}$ in configuration space and the spasing state was determined as the one whose eigenvalue $\Lambda_{s}(\omega)$ has the largest imaginary part $\Lambda_{s}''(\omega)$. Note that the three-fold degeneracy (for $l=1$) of a spherical NP is broken down by random distribution of gain molecules in the shell so there are no degenerate eigenvalues. The resonance frequency $\omega_{s}$ and the threshold value $n_{s}$ were determined by solving Eqs. (\ref{resonance2}) and (\ref{loss_condition}), respectively. To distinguish between quenching and Coulomb effects, we compare the results for dipole mode only $l=1$ with those for up to $l=50$ terms in the matrix $D_{jk}^{p}$ calculated, in both cases, with and without the direct dipole-dipole coupling term $D_{jk}^{d}$.

 \begin{figure}[bt]
 \centering
 \includegraphics[width=1\columnwidth]{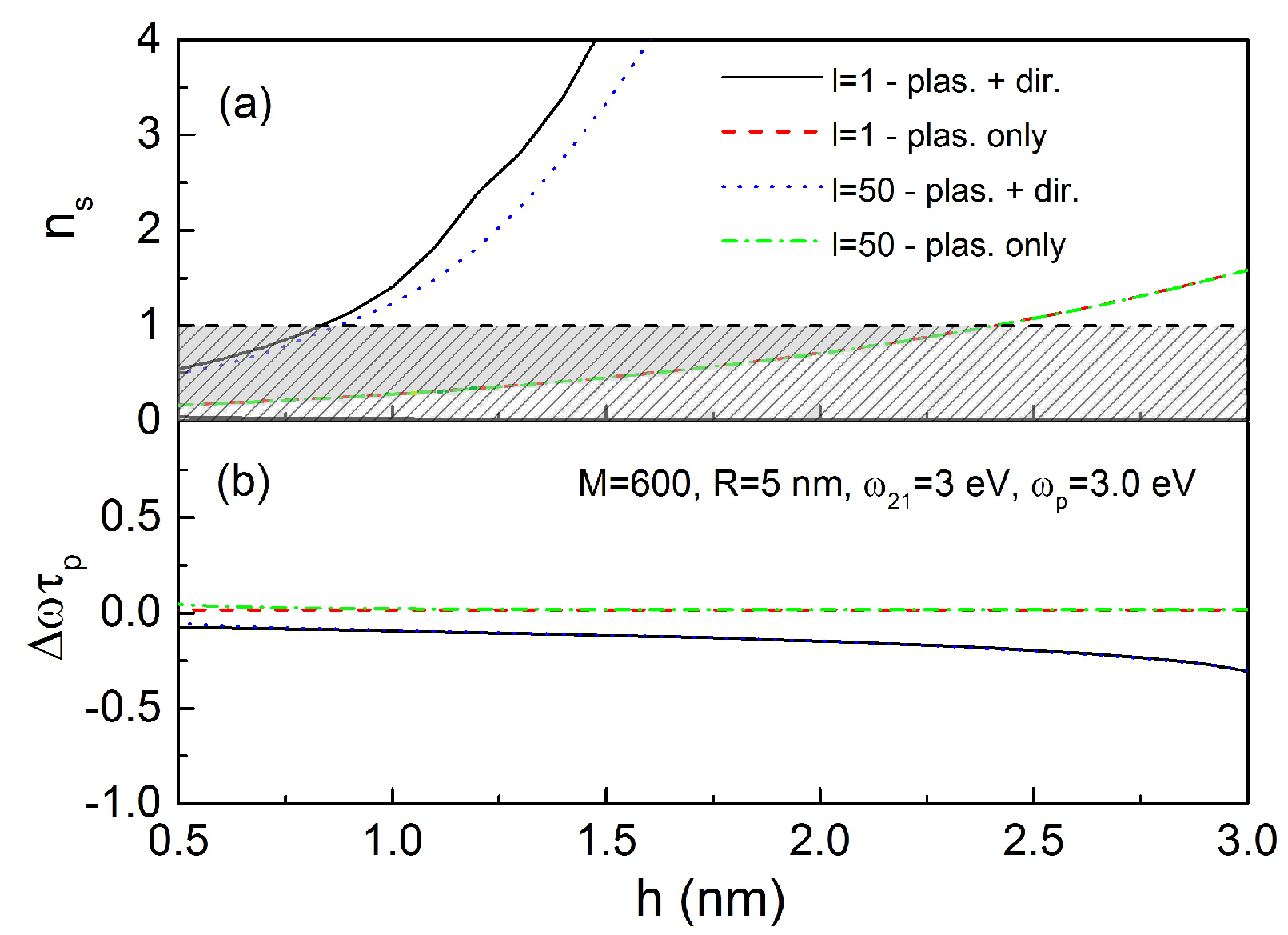}
 \caption{\label{fig:M600} (a) Spasing threshold $n_{s}$, the hatched region represents the gain condition and the spasing region is shaded grey, and (b)  frequency shift $\Delta \omega=\omega_{s}-\omega_{21}$ for $M=600$ molecules with $\omega_{21}=\omega_{p}$ are plotted vs. shell thickness $h$   with and without direct coupling for dipole ($l=1$) plasmon mode and for up to $l=50$ modes included.}
  \end{figure}

In Fig. \ref{fig:M600} we show resonance frequency shift $\Delta\omega=\omega_{s}-\omega_{21}$, normalized by plasmon lifetime $\tau_{p}$, and threshold population inversion per molecule $n_{s}=N_{s}/M$ as a function of shell thickness $h$ for $M=600$ gain molecules randomly distributed in the shell on top of $R_{c}=5$ nm Ag core. The gain frequency $\omega_{21}$ was chosen to coincide with the dipole plasmon frequency $\omega_{p}\approx 3.0$ eV for the parameters chosen. In the single mode case ($l=1$) and in the absence of direct dipole-dipole coupling, the calculated $\Delta\omega$ is nearly vanishing, in agreement with Eq. (\ref{resonance3}), while $n_{s}$ increases with $h$ before reaching its maximum value $n_{s}=1$ at $h\approx 2.35$ nm. This threshold behavior is consistent with the condition (\ref{condition}) as the latter implies the increase of $N$ with mode volume until the full population inversion $N=M$ is reached which, in the case of low gain molecule number $M=600$, takes place for relatively small shell thickness.

Very similar results are obtained when higher $l$ modes (up to $l=50$) are incorporated in the coupling matrix $D_{jk}^{p}$ in Eq.\ (\ref{matrix1}). Neither $\Delta\omega$ nor $n_{s}$ show significant deviations from the  $l=1$ curves except for unrealistically small shell thickness below $0.5$ nm (not shown here). This behavior should be contrasted to the single molecule case, where the molecule decay into high $l$ modes leads to  fluorescence quenching at several nm distances from the NP surface \cite{silbey-acp78,metiu-pss84,feldmann-prl02,lakowicz-jf02,artemyev-nl02,gueroui-prl04,feldmann-nl05,klimov-jacs06,strose-jacs06,mertens-nl06,pompa-naturenanotech06,novotny-prl06,sandoghdar-prl06,novotny-oe07,sandoghdar-nl07,lakowicz-jacs07,chen-nl07,lakowicz-nl07,halas-nl07,feldmann-nl08,halas-acsnano09,ming-nl09,kinkhabwala-naturephot09,viste-acsnano10,lakowicz-jacs10,munechika-nl10,ming-nl11,ratchford-nl11,raino-acsnano11}. A similar quenching effect was demonstrated in cooperative emission of relatively small number  ($M<100$) of dyes \cite{pustovit-prl09,pustovit-prb10}. For larger ensembles, however, the quenching effects apparently become insignificant  due to the effective restoration of spherical symmetry that inhibits the mode mixing.

Turning the direct dipole-dipole interactions between gain molecules to a maximum, described by the matrix $D_{jk}^{d}$ in Eq.\ (\ref{matrix1}), has dramatic effect both on resonance frequency and threshold population inversion. The resonance frequency exhibits negative shift relative to the plasmon frequency whose amplitude  increases with  $h$. The overall negative sign of $\Delta \omega$ is due to the normal orientation of molecule dipoles relative to NP surface, while the increase of $|\Delta \omega|$ with $h$ is due to reduced plasmonic contribution   $D_{jk}^{p}$ which has the opposite sign and decreases with $h$ faster than the direct contribution $D_{jk}^{d}$. Note that real systems would lie somewhere between the non-interacting case and this maximum dipole-dipole interaction case where the choice of molecules' normal dipole orientation may overestimate $|\Delta \omega|$ as compared to a more realistic random orientations. Even so, the new resonance frequency lies well within the plasmon spectral band (i.e., $\tau_{p} \Delta\omega  \ll 1$). At the same time, the maximal threshold value $n_{s}=N_{s}/M=1$ is reached at about $h=1$ nm, indicating that, in the presence of direct coupling between gain molecules, the dependence (\ref{condition}) is no longer valid. Note that here the mode mixing has somewhat larger effect than in absence of direct coupling presumably due to the violation of spherical symmetry by  much stronger interactions between closely spaced molecules.

 \begin{figure}[bt]
 \centering
 \includegraphics[width=1\columnwidth]{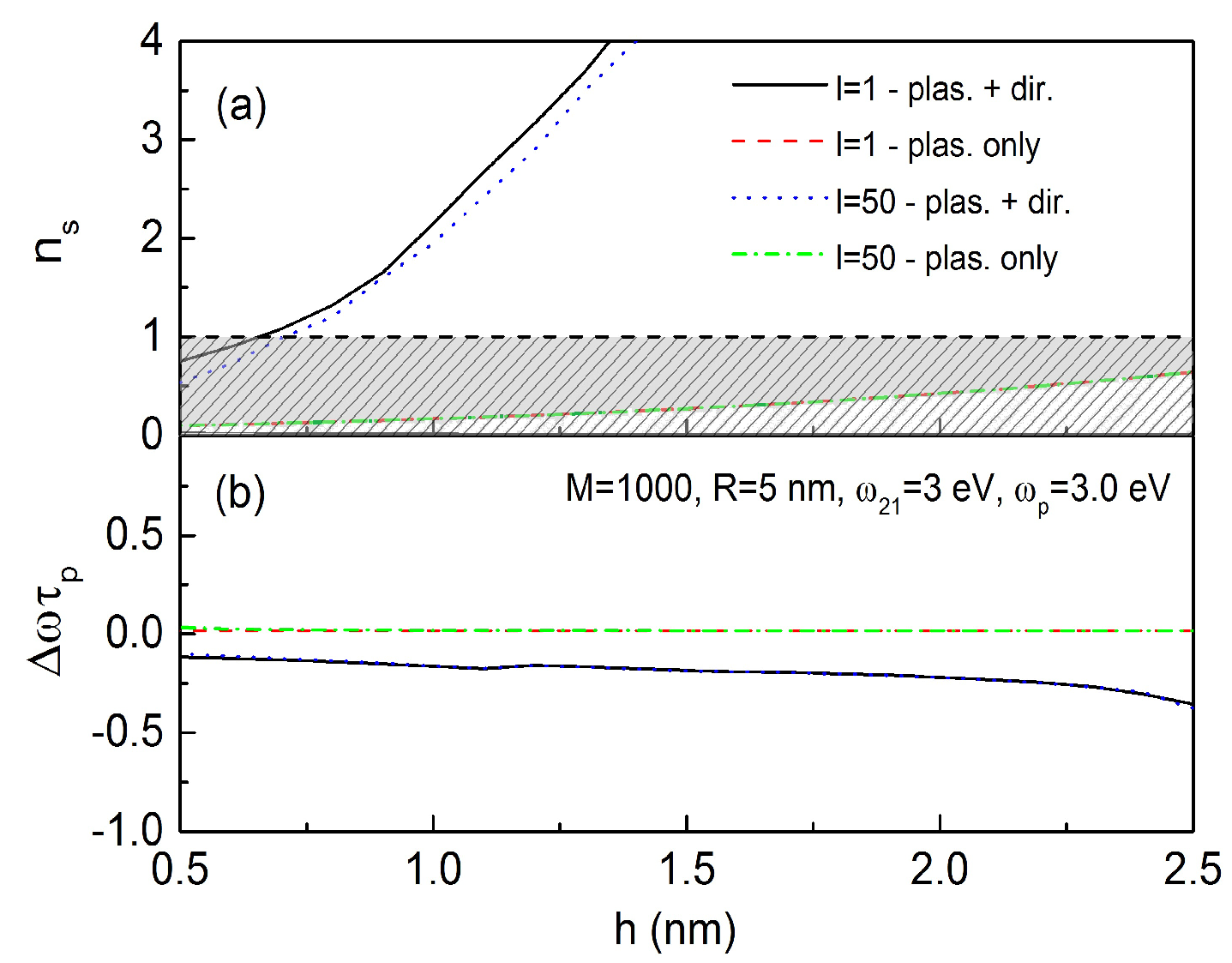}
 \caption{\label{fig:M1000} (a) Spasing threshold $n_{s}$, the hatched region represents the gain condition and the spasing region is shaded grey, and (b)  frequency shift $\Delta \omega=\omega_{s}-\omega_{21}$ for $M=1000$ molecules with $\omega_{21}=\omega_{p}$ are plotted vs. shell thickness $h$   with and without direct coupling for dipole ($l=1$) plasmon mode and for up to $l=50$ modes included.}
  \end{figure}

In Fig. \ref{fig:M1000}, we repeat  our calculations for a larger number of gain molecules, $M=1000$, that show two notable differences with the $M=600$ case. In the absence of direct coupling between gain molecules, the maximal threshold value $n_{s}=1$ is reached at larger shell thickness values, in agreement with  Eq. (\ref{condition}). However, when the direct coupling is turned on, the maximal threshold value  is reached at \textit{smaller} value of $h\approx  0.75$ nm, which must be attributed to  stronger dipole-dipole interactions  for higher gain densities. At the same time, the effect of mode mixing in $n_{s}$ dependence on $h$ becomes more pronounced, which is also related  to stronger interactions between more closely spaced molecules that can effectively break spherical symmetry in a larger system.

 \begin{figure}[tb]
 \centering
 \includegraphics[width=1\columnwidth]{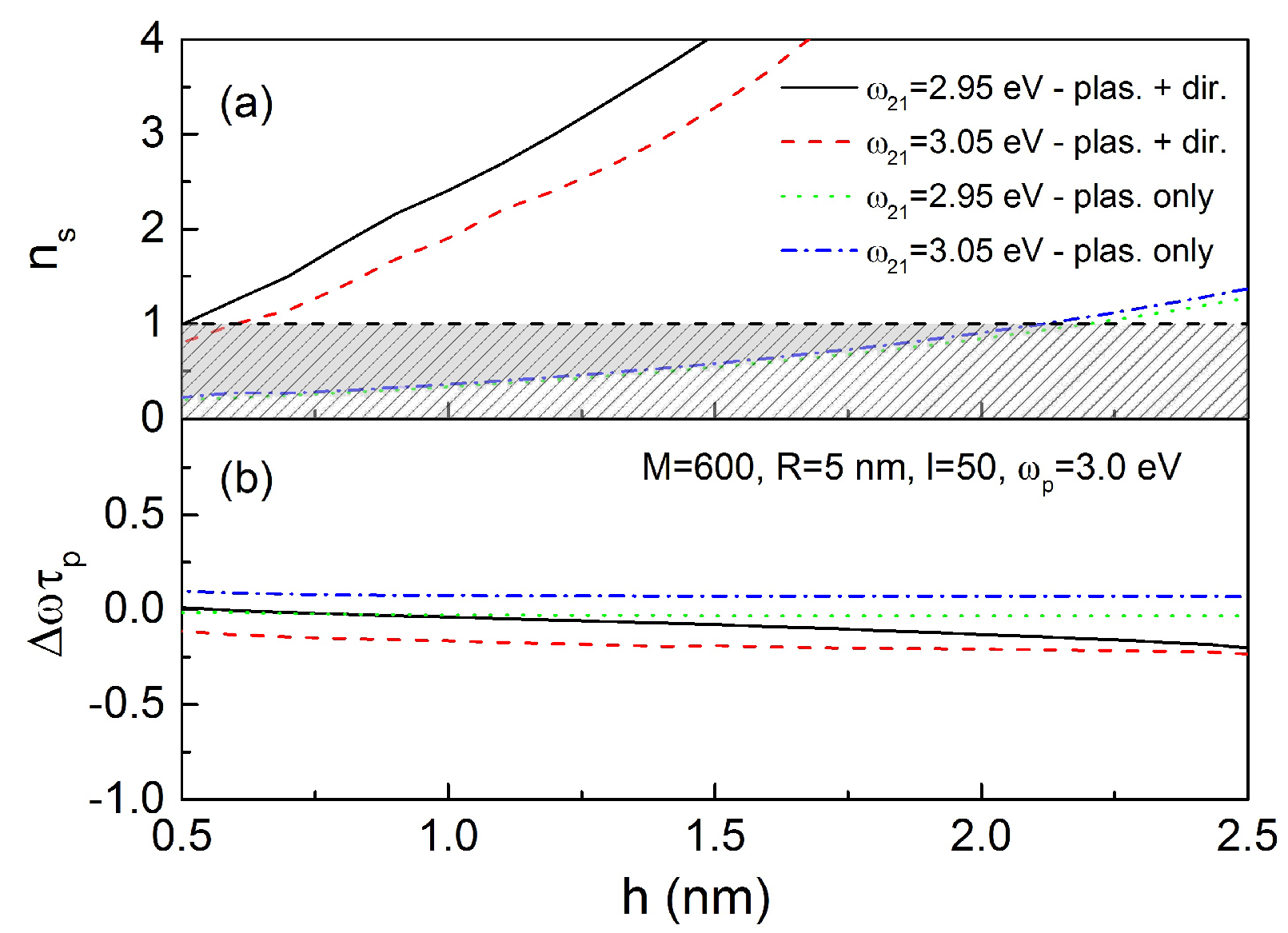}
 \caption{\label{fig:M500-shift} (a) Spasing threshold $n_{s}$, the hatched region represents the gain condition and the spasing region is shaded grey, and (b)  frequency shift $\Delta \omega=\omega_{s}-\omega_{21}$ for $M=600$ molecules are plotted vs. shell thickness $h$  for gain spectral bands centered at 2.95 eV and 3.05 eV.}
  \end{figure}

 \begin{figure}[tb]
 \centering
 \includegraphics[width=1\columnwidth]{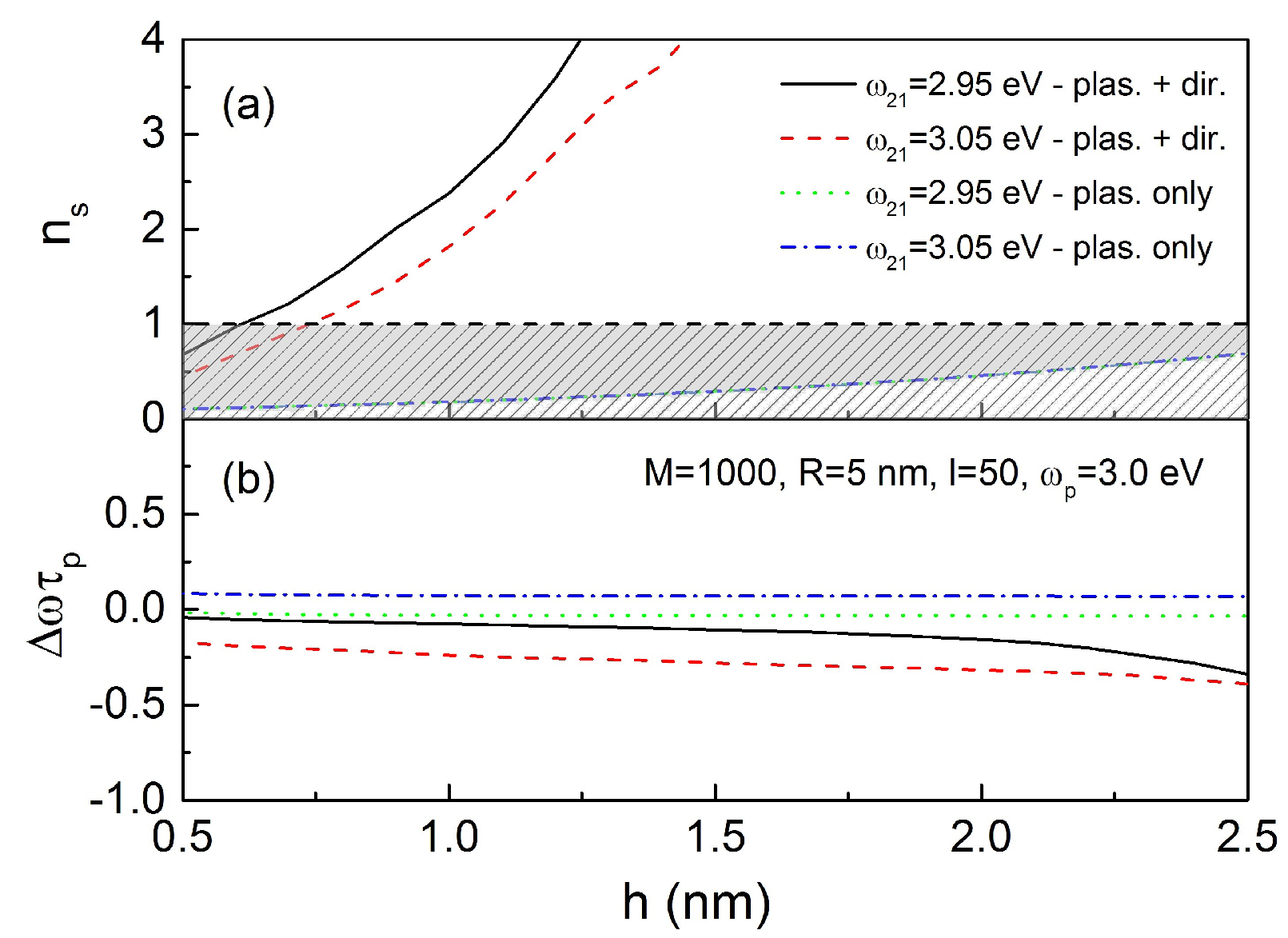}
 \caption{\label{fig:M1000-shift} (a) Spasing threshold $n_{s}$, the hatched region represents the gain condition and the spasing region is shaded grey, and (b)  frequency shift $\Delta \omega=\omega_{s}-\omega_{21}$ for $M=1000$ molecules are plotted vs. shell thickness $h$  for gain spectral bands centered at 2.95 eV and 3.05 eV.}
  \end{figure}

The major effect of direct dipole-dipole interactions  is the \textit{random} Coulomb shift of gain molecules' excitation frequencies which may lead to the detuning between  individual gain molecules and SP resonance. Note that the \textit{average} negative shift that is due to normal orientation of molecular dipoles can be compensated by changing the gain molecules' excitation frequency. In  Figs.\ \ref{fig:M500-shift} and \ref{fig:M1000-shift} we show calculated $\Delta\omega$ and $n_{s}$  both for redshifted ($\omega_{21}=2.95$ eV) and blueshifted ($\omega_{21}=3.05$ eV) gain frequencies relative to the SP resonance at 3.0 eV. As expected, for $\omega_{0}=2.95$ eV, the average shift of $\Delta\omega$ is strongly reduced while it increases for $\omega_{0}=3.05$ eV [see Fig.\ \ref{fig:M500-shift}(b) and  Fig.\ \ref{fig:M1000-shift}(b)]. However, the maximal threshold value $n_{s}=1$   is now reached for even smaller shell thickness $h< 0.5$ nm [see Fig.\ \ref{fig:M500-shift}(a) and  Fig.\ \ref{fig:M1000-shift}(a)], indicating that the loss of coherence is caused by the \textit{fluctuations} of gain excitation energies.

\begin{figure}[tb]
 \centering
 \includegraphics[width=1\columnwidth]{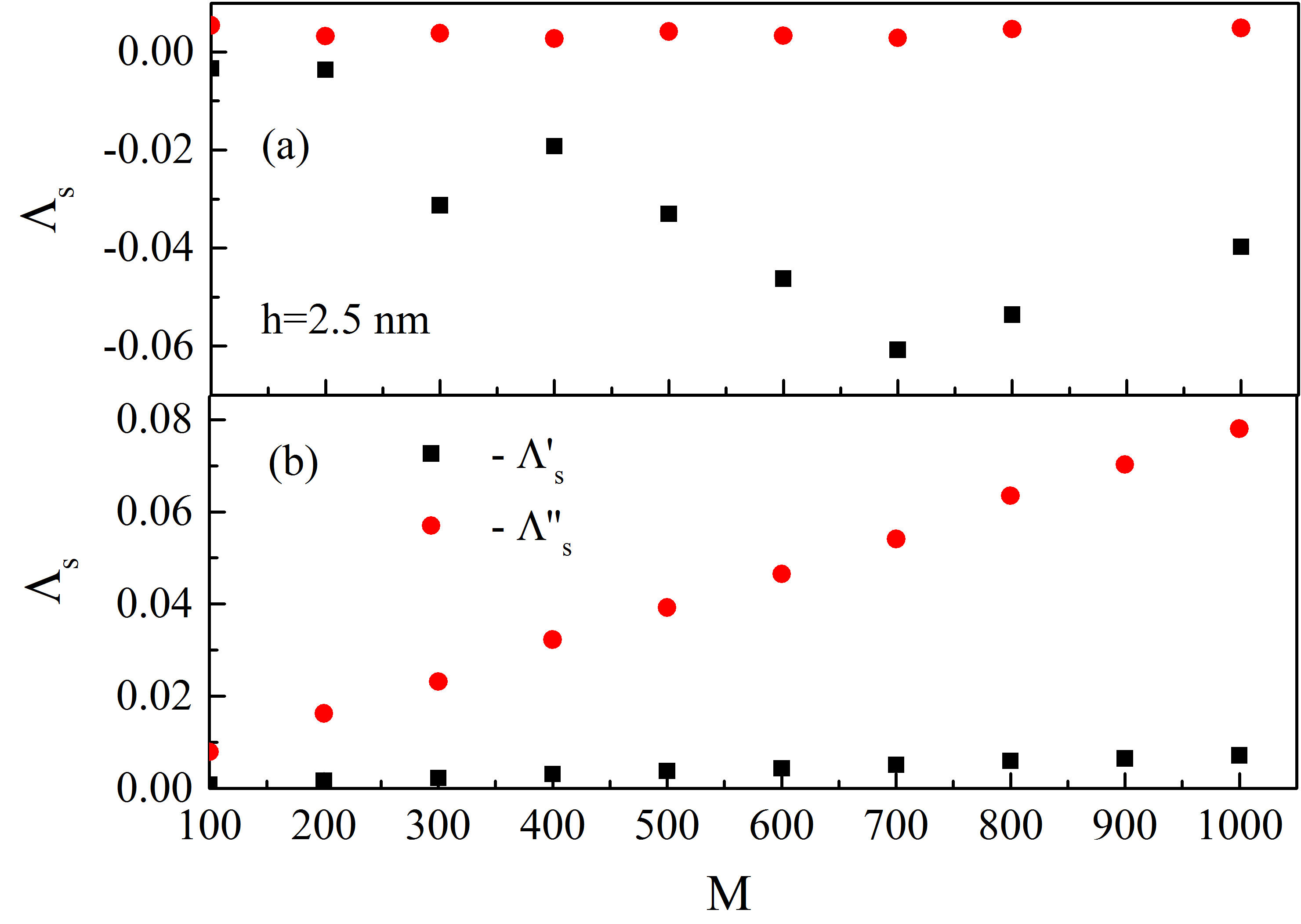}
 \caption{\label{fig:eigen-M} Calculated real $\Lambda'_{s}$ and imaginary $\Lambda''_{s}$ parts of eigenvalue $\Lambda_{s}$  are shown for different gain molecule numbers $M$  (a) with  and (b) without direct coupling. The thickness of gain layer is set to $h=2.5$ nm.}
  \end{figure}

To pinpoint the loss of coherence, we show in Fig.\ \ref{fig:eigen-M} the  calculated eigenvalues $\Lambda_{s}$  for different gain molecule numbers $M$ both with and without dipole-dipole interactions. According  to Eq. (\ref{coop}), the coherence implies that $\Lambda_{s}$  scales \textit{linearly} with $M$ to ensure that the condition (\ref{condition1}) is size-independent (for constant density of inverted molecules, $N/V_{m}$). This is indeed the case in the \textit{absence} of direct interactions between gain molecules: both real and imaginary parts of $\Lambda_{s}$ scale nearly linearly with  $M$ ranging from 100 to 1000. However, with direct coupling turned on, neither of them shows linear dependence on $M$, implying that the condition (\ref{condition}) no longer holds.  Instead, $\Lambda''_{s}$ is nearly constant while $\Lambda'_{s}$ shows large fluctuations,  especially for larger values of $M$, presumably due to larger  frequency shifts at higher densities.

Finally, we note that real systems would lie somewhere between the two extreme states of molecules dipole orientations normal to NP surface ( likely overestimating the role of Coulomb shifts of gain excitation frequencies) and the non-interacting case as the random dipole orientations in actual NP-based spasers likely weakens the negative effect of  direct interactions  on spasing threshold. Note also that, for larger systems, the fluctuations of gain excitation frequencies are expected to be weaker. Our numerical results are not sufficient to establish a new threshold condition that would replace Eq. (\ref{condition}) in small systems.  Nevertheless, our calculations indicate that the direct interactions identify a parameter window in which spasing threshold can be realistically achieved.

\section{Conclusions}
\label{sec:conc}
In summary, we performed a numerical study of the effect of mode mixing and direct dipole-dipole interactions between gain molecules on spasing threshold for small composite nanoparticles with metallic core and dye-doped dielectric shell. We found that for sufficiently large ($\sim 1000$)  gain molecule numbers, the quenching is negligibly small and a single-mode approximation should work well for realistic systems. In contrast, we found that direct dipole-dipole interactions, by causing random Coulomb shifts of gain molecules' excitation frequencies, may lead to system dephasing and hinder reaching the spasing threshold in small systems. These two regimes serve as edges to an identified parameter window in which spasing can likely be achieved.

\section{Acknowledgments}
This research was performed while the first author held a National Research Council Research Associateship Award at Air Force Research Laboratory. This work was also supported by AFRL Materials and Manufacturing Directorate Applied Metamaterials Program. Work in JSU was supported in part by the NSF under grant No. DMR-1206975.


%
\end{document}